\documentclass[twocolumn,aps,prd,superscriptaddress,preprintnumbers,nofootinbib,nobalancelastpage]{revtex4-2}

\usepackage{mathrsfs} 
\usepackage{amsfonts}
\usepackage{amsmath}
\usepackage{amsthm}
\usepackage{amssymb}
\usepackage{array}
\usepackage{dcolumn}
\usepackage[normalem]{ulem}
\usepackage[svgnames]{xcolor}
\usepackage{hyperref}
\usepackage{balance}
\hypersetup{
    colorlinks=true,
    linkcolor=NavyBlue,
    filecolor=NavyBlue,
    urlcolor=NavyBlue,
    citecolor=NavyBlue,
    }
\usepackage{graphics}
\usepackage{graphicx}
\usepackage[caption=false]{subfig}
\usepackage{adjustbox}
\usepackage{breqn}
\usepackage{fontawesome} 
\usepackage{slashed}

\def\be{\begin{equation}}
\def\ee{\end{equation}}

\def\bea{\begin{eqnarray}}
\def\eea{\end{eqnarray}}

\newcount\Comments  
\newcommand{\kibitz}[2]{\ifnum\Comments=1\textcolor{#1}{#2}\fi}

\makeatletter
\let\cat@comma@active\@empty
\makeatother

\begin{document}

\title
{The Penrose Transform and the Kerr-Schild double copy}

\preprint{LA-UR-25-31238}

\author{Emma Albertini}
\affiliation{SISSA, Via Bonomea 265, 34136 Trieste, Italy \& INFN Sezione di Trieste} \affiliation{IFPU - Institute for Fundamental Physics of the Universe, Via Beirut 2, 34014 Trieste, Italy}
\affiliation{Blackett Laboratory, Imperial College London, SW7 2AZ, U.K}

\author{Michael L. Graesser}
\affiliation{Theoretical Division, Los Alamos National Laboratory, Los Alamos, NM 87545, USA}

\author{Gabriel Herczeg}
\affiliation{Brown Theoretical Physics Center, Department of Physics, Brown University, Providence, RI, 02912, USA}

\begin{abstract}
There are a number of classical double copies, each providing a prescription for generating solutions to the Maxwell and scalar wave equations from exact solutions of Einstein's equations.  Two such prescriptions are the Kerr-Schild and twistorial double copies. 
We argue that for a broad class of self-dual vacuum solutions of the Kerr-Schild form, which we refer to as {\em twistorial Kerr-Schild spacetimes}, these two prescriptions are in fact equivalent. The approach is elementary, utilizing null Lorentz transformations, with homogeneous functions on twistor space playing a central role. 
The equivalence is illustrated explicitly for the example of the self-dual (Kerr)-Taub-NUT spacetime. A detailed proof and several more examples will be presented in a long-form companion to this letter. 
\end{abstract}

\maketitle

{\em Introduction.}---
There now exists a remarkable web of relations between gravitational and gauge field theories. Ample evidence of the deep connections between gravitational and gauge theories is provided by the Kawai-Lewellen-Tye `squaring relations' for open and closed string scattering \cite{Kawai:1985xq}, holography \cite{tHooft:1993dmi, Susskind:1994vu}, 
the AdS/CFT correspondence \cite{Maldacena:1997re, Gubser:1998bc,Witten:1998qj},
the Bern-Carrasco-Johansson (BCJ) colour-kinematics duality for gluon/graviton scattering in field theory \cite{Bern:2008qj,Bern:2010ue,Bern:2010yg,Bern:2013yya}, and continuing developments in flat-space holography \cite{Strominger:2017zoo}. These advances in gauge/gravities dualities over the last few decades provide theoretical context for the classical double copy \cite{Monteiro:2014cda, Luna:2015paa, Luna:2018dpt}, which is the focus of this work. Recent reviews 
can be found in Refs. \cite{Borsten:2020bgv,Bern:2019prr, White:2021gvv, Bern:2022wqg, Adamo:2022dcm}.

Among the many families of exact solutions to the Einstein field equations \cite{Stephani:2003tm}, one class of solutions stands out in the historical development of the classical double copy. These are the Kerr-Schild metrics, among which,  
the real vacuum Kerr-Schild spacetimes are algebraically special with Petrov type II, D, or N \cite{Chandrasekhar:1985kt}, and include the Kerr metric as a distinguished example.

A Kerr-Schild spacetime provides the raw elements for the classical (Kerr-Schild) double copy \cite{Monteiro:2014cda, Luna:2015paa, Luna:2016due}: from the metric data one constructs a harmonic function---the `zeroth copy'---as well as a solution to the source-free Maxwell equations on Minkowski spacetime---the `single copy'. An instance of the classical double copy is the Weyl double copy \cite{Luna:2018dpt}, which uses the Weyl tensor to form a solution to the Maxwell equations. There is also the Newman-Penrose map \cite{Elor:2020nqe, Farnsworth:2021wvs, Farnsworth:2023mff}, extensions of the Kerr-Schild double copy to curved backgrounds \cite{Bahjat-Abbas:2017htu, Carrillo-Gonzalez:2017iyj}, among many others \cite{Garcia-Compean:2024zze, Beetar:2024ptv, Chawla:2025uwu, Chawla:2024mse, Holton:2025slh, Holton:2025hny, Kent:2025pvu, Caceres:2025eky, Chacon:2024qsq, Zhao:2024wtn, Zhao:2024ljb, Armstrong-Williams:2024bog, LopesCardoso:2024ttc, CarrilloGonzalez:2024sto, Ilderton:2024oly, Liu:2024byr, Garcia-Compean:2024uie, Garcia-Compean:2024zze, Mao:2023yle, Chen:2023kjo, Ortaggio:2023cdz, Liang:2023zxo, Brown:2023zxm, Borsten:2023paw, Diaz-Jaramillo:2025gxw, Bonezzi:2023pox, Lipstein:2023pih, CarrilloGonzalez:2022ggn, Nagy:2022xxs, Goldberger:2016iau, Adamo:2017nia, Berman:2018hwd, Lee:2018gxc, Alawadhi:2019urr, Huang:2019cja, Arkani-Hamed:2019ymq, Lescano:2020nve, Lescano:2021ooe, Rodriguez:2025hiu, Campiglia:2021srh, Adamo:2021dfg, Godazgar:2021iae, 
Carrasco:2025bgu, Emond:2025nxa, Moynihan:2025vcs, Ilderton:2025aql, Aoude:2025jvt}.

A subject of the present work is the twistorial version of the Weyl double copy \cite{White:2020sfn, Chacon:2021wbr, Chacon:2021lox}. This approach 
follows Penrose's construction of massless spin fields through use of 
a contour integration 
of meromorphic functions on projective twistor space, using the incidence relation \cite{Penrose:1965am, Penrose:1967wn, Penrose:1968me}. 

Here we initiate a study of the relations between these double copies. We argue that for a broad class of exact, self-dual solutions of the vacuum Einstein equations, the twistorial and Kerr-Schild double copies are equivalent. The key ingredient allowing us to establish this equivalence is the fact that these solutions are essentially completely determined by homogeneous functions on twistor space, which can be inputed directly into the Penrose transform to obtain solutions of the zero rest mass equation with any spin. At spin two, this leads to a solution of the \emph{linearized} Einstein equations---however, the Kerr-Schild ansatz already linearizes the Einstein equations, and as such, vacuum Kerr-Schild spacetimes are simultaneously solutions of linearized gravity and of the full non-linear Einstein equations. 

{\em The Kerr-Schild Double Copy.}---A Kerr-Schild spacetime is described by a metric of the form
\begin{align}
    g_{\mu \nu} = \eta_{\mu \nu} + 2 V \ell_\mu \ell_ \nu 
    \label{eq:KSmetric}
\end{align}
where  $\ell$ is assumed to be null. A real vacuum spacetime of this form is necessarily algebraically special, and $\ell$ is a repeated principal null direction \cite{Chandrasekhar:1985kt}. 
The Goldberg-Sachs theorem 
asserts the existence of a geodetic, shear free, null congruence if and only if the spacetime is algebraically special \cite{2009GReGr..41..433G}. Thus for real spacetimes,  $\ell$ is not only null, but also shear-free and geodetic. 

The extension of spacetimes into the complex domain generalizes real, Lorentzian signature geometries into a unified framework that also includes Euclidean and Kleinian signature metric as special cases. Moreover, there are no real, Lorentzian self-dual metrics in four spacetime dimensions, so it is natural to consider complexified spacetimes, as well as real Kleinian and Euclidean signature metrics in the self-dual setting. 
Such self-dual configurations have long been studied in the context of twistor theory, integrable systems, and Euclidean quantum gravity \cite{Witten1978, GibbonsHawking1978, Atiyah1978}, and were studied in the context of the double copy in \cite{Monteiro:2020plf}. Kleinian signature allows for nontrivial on-shell three-point amplitudes—objects that vanish in Lorentzian signature—which can be used to recursively construct higher-point scattering amplitudes. Consequently, classical solutions of Einstein’s equations with Kleinian signature have attracted renewed attention. Moreover, in light of asymptotic symmetries and holography,
the self-dual sectors of gravity can be viewed as analytically tractable corners of the full gravitational phase space where the asymptotic structure simplifies while preserving nontrivial topology and curvature \cite{Crawley_2022,Crawley_2023}.

In the Newman-Penrose formalism for general relativity \cite{Newman:1961qr}, 
a null tetrad basis 
is introduced, with complex spin-coefficients encoding the information of the linear connection. This formalism reduces the second-order Einstein equations to first order, at the expense of introducing more variables (the spin-coefficients) and equations governing their dynamics. Despite the formidable complexity of the equations, this formalism has greatly facilitated solving the vacuum Einstein and Einstein-Maxwell equations in the algebraically special setting. Here the Kerr-Schild metric (\ref{eq:KSmetric}) is given by 
\bea 
g &=& 2(\theta^1  \theta^2 - \theta^3 \theta^4) \nonumber
\eea
where $\theta^1_\mu= n_\mu$, $\theta^2_\mu= \ell_\mu$, $\theta^3_\mu= - \tilde{m}_\mu$, and $\theta^4_\mu= -m_\mu$, are the four null vector fields of the tetrad frame satisfying the orthonormality conditions $\ell \cdot n=1=-m \cdot \tilde{m}$, with all other inner products vanishing. A co-tetrad compatible with the Kerr-Schild form is provided by 
\begin{eqnarray}
\theta^1 &=&dv+ V \theta^2, \quad
\theta^2 = du+Y d\tilde{\zeta} + \tilde{Y} d \zeta + Y \tilde{Y} dv, \nonumber \\ \theta^3&=& d \tilde{\zeta} + \tilde{Y} dv, \quad
\theta^4 = d \zeta + Y dv \, . \nonumber
\end{eqnarray}
In four complex spacetime dimensions $u,v, \zeta, \tilde{\zeta}$ are null coordinates, and $Y$, $\tilde{Y}$, and $V$ are complex functions of these coordinates. The directional derivatives are defined by
$D=\ell^\mu \partial_\mu$, $\Delta =n^\mu \partial_\mu$, $\delta=m^\mu \partial_\mu$, $\tilde{\delta}=\tilde{m}^\mu \partial_u$, 
In terms of this tetrad they are given by 
\begin{eqnarray}
D &=& \partial_v- Y \partial_\zeta - \tilde{Y} \partial_{\tilde{\zeta}} + Y \tilde{Y} \partial _u, \quad \Delta = \partial_u - V D, \nonumber \\
\delta &=&  \partial_{\tilde{\zeta}}-Y \partial_u, \quad \tilde{\delta}=\partial_\zeta-\tilde{Y} \partial_u. 
\end{eqnarray}
For complex spacetimes $Y$ and $\tilde{Y}$ are in general independent of each other, 
whereas for real spacetimes $\tilde{Y}=Y^*$, with $Y$ complex.
For real spacetimes, $u$, $v$ are real, $\tilde{\zeta}=\zeta^*$ is complex. Within the complex setting, we may select $Y \neq 0$, and $\tilde{Y} = 0$, which leads to a self-dual solution. In the self-dual case, it is often convenient to regard $\zeta$ and $\tilde{\zeta}$ as independent \emph{real} coordinates, in which case, the spacetime has Kleinian signature.

The general solution for real vacuum Kerr-Schild spacetimes of Minkowskian signature that are {\em expanding} was fully characterized by Debney, Kerr, and Schild, whose results we state here before moving on to self-dual spacetimes \cite{Debney:1969zz}:

{\em A real vacuum Kerr-Schild spacetime of Minkowskian signature is algebraically special, with the null vector $\ell$ being a repeated principal null direction of the Weyl tensor, and hence, geodetic and shear-free. The solution $Y$ to the vacuum gravitational field equations is given by the zero set of the function 
\begin{gather}
f= \Phi(Y)+ (\tilde{c} Y + a)(\zeta + Y v) - 
(bY + c)(u + Y \tilde{\zeta})
\label{eqn:F}
\end{gather}
for constants $a, b \in \mathbb{R}$, $c \in \mathbb{C}$ and $\tilde{c}=c^*$. Here $\Phi(Y)$ is an arbitrary local analytic function of $Y$. 
Moreover, the vacuum Kerr-Schild spacetime admits the Killing vector 
\begin{gather}
    X= a\partial_u + b\partial_v + c\partial_{\zeta} + \tilde{c}\partial_{\tilde{\zeta}} \,.
    \label{eqn:Killing-vector}
\end{gather}
The function $V$ is real and given by $V=m (\rho + \rho^*)/2P^3$, for an arbitrary real constant $m$, and where $P$ is the real function $P=a + \tilde{c} Y+ c \tilde{Y} + b Y \tilde{Y}$.} We note a metric is {\em expanding} when $Re(\rho) \neq 0$.

The set of such spacetimes is evidently rather large.
Included in this set are the physically important Schwarzschild and Kerr black hole solutions. While we focus exclusively on vacuum solutions here, it should be noted that this approach can be generalized straightforwardly to the Einstein-Maxwell system, which can incorporate the Reissner-N\"{o}rdstrom and Kerr-Newman charged black hole solutions \cite{Stephani:2003tm}. Solutions of this type have been studied in the context of the Weyl double copy in \cite{Easson:2021asd, Easson:2022zoh}.

The Kerr-Schild double copy is implemented in the Newman-Penrose formalism in the following way. On the Kerr-Schild side, the zeroth copy is defined by $\varphi^{KS} = h_{ab}X^a X^b$, where $h_{ab} = 2Vl_a l_b$ is the Kerr-Schild graviton. The scalar $\varphi^{KS}$ is harmonic on flat-space \cite{Easson:2023dbk}. The single copy is given by  $A_\mu = h_{\mu \nu} X^v$. The information of the field strength $F=dA$ is encoded in the self-dual Maxwell scalars $\varphi_i$ given by \cite{Easson:2023dbk}
\begin{gather*}
\begin{split}
\varphi_0=0, \quad \varphi_1 = 2 P \gamma, \quad
\varphi_2=2(P\nu+ V \tilde{\delta}P) \, .
\end{split}
\end{gather*}
In the self-dual limit, the spin-coefficients are $\gamma=m\rho^2/2P^3$, and $\nu = \tilde{\delta} V$. One can verify that $F$ satisfies the source-free Maxwell equations of the Kerr-Schild and flat-space backgrounds \cite{Easson:2023dbk}.
Finally, the Weyl scalars, which encode all the information in the Weyl tensor, are given by 
\begin{align*}
\begin{split}
\Psi_0=\Psi_1 =0 &, \quad 
\Psi_2 
= \frac{m \rho^3}{P^3}, \quad
\Psi_3=\frac{3}{2} \frac{m \rho^2}{P^3}\left(\frac{\tilde{\delta} \rho}{\rho}+ 2\frac{\rho}{\tilde{\rho}}\tilde{\tau}\right),  \\
\Psi_4 &=\frac{m}{2P^3}\left(\tilde{\delta}^2 \rho+ 9 \frac{\rho}{\tilde{\rho}} \tilde{\tau} 
     \tilde{\delta} \rho+ 12 \frac{\rho^3}{\tilde{\rho}^2} \tilde{\tau}^2 \right)
     \quad .
 \end{split}
 \nonumber
\end{align*}

{\em Twistorial Kerr-Schild spacetimes.}---The full characterization of all real vacuum Kerr-Schild spacetimes by Debney, Kerr, and Schild generalizes only in part to complex spacetimes, however. Complex vacuum Kerr-Schild spacetimes are not necessarily algebraically special, 
nor always admitting a Killing vector, 
nor are $Y$ and $\tilde{Y}$ necessarily algebraic. 
 
That said, 
the holomorphic function $f$ in (\ref{eqn:F}), 
whose zero set generates a $Y$, and an independently specified anti-holomorphic function $\tilde{f}$ of an analogous form, whose zero set gives a  $\tilde{Y}$, along with $V$ and $P$ of the above form, does describe a family of metrics that satisfy the vacuum Einstein equations. In general, such solutions are complex. 

In what follows we will {\em assume} a complex vacuum Kerr-Schild solution of this special form, which we refer to as {\em twistorial Kerr-Schild metrics}, for reasons that will soon be clear. They have been extensively studied in, for example \cite{mcintosh1989kerr,mcintosh1988single,mcintosh-complex-relativity-I,Easson:2023dbk}, whose results we make use of.

To construct self-dual solutions, 
the main observation is that given a homogeneous and meromorphic function $K(Z^\alpha)$ on twistor space, the zero set of $K$ provides a geodetic, and shear-free null congruence 
$Y$ on Minkowski space.

This statement, often referred to as Kerr's Theorem, first appeared in a publication by Penrose, although Penrose attributes the discovery of the theorem to Kerr \cite{Penrose:1967wn}. While Penrose used twistor theory to establish this result, a qualitatively different proof of Kerr's theorem, assuming only that $f(Y,X_1,X_2)$ is meromorphic, was later given by Cox and Flaherty using the Newman-Penrose formalism and integrability conditions \cite{1976CMaPh..47...75C}. 

A function $f(Y,X_1,X_2)$ can be obtained from $K$ by dehomogenizing to the affine space $Z^2=1$, 
with  $Z^0= X_1 Z^2 $,
$Z^1=iX_2 Z^2$, 
and $Y \equiv Z^3/Z^2$. 
To make contact with previous notation, the incidence relation 
described further below, is used to relate $X_1$ and $X_2$ to the spacetime coordinates and $Y$: $X_1 = \zeta + Y v$ and $X_2=u + Y \tilde{\zeta}$. 
Finally, any meromorphic $f(Y,X_1,X_2)$ can be lifted to a homogeneous meromorphic function $K$ on twistor space.

As Debney, Kerr, and Schild showed \cite{Debney:1969zz} for real vacuum Kerr-Schild spacetimes, the functional form of $f$ is further constrained by the requirement that the metric induced by a zero set of $f$ satisfies the Einstein field equations. In particular, for this class of spacetimes $f$ has
the specific form 
\begin{gather}
f(Y,X_1,X_2)= \Phi(Y) + (\tilde{c} Y + a)X_1 - 
(bY + c)X_2 \qquad . \nonumber
\end{gather}
A self-dual vacuum Kerr-Schild solution is then constructed from $Y$ and $P$, where $V=m\rho/2P^3$, 
$P = l_a X^a$,
and $\tilde{f}(\tilde{Y},\tilde{X}_1,\tilde{X}_2)=\tilde{Y}$. This last condition trivially implies $\tilde{Y}=0$. Henceforth we restrict our discussion to self-dual solutions of this form. 

A null tetrad for a given metric is not unique, for there exists the freedom to perform local Lorentz transformations which rotate the tetrad while leaving the metric invariant. For real spacetimes these so-called ``null rotations" can be described by the covering group $SL(2, \mathbb{C})$, whereas for complex spacetimes the covering group for null rotations is given by $SL(2, \mathbb{C})_L \times SL(2,\mathbb{C})_R$. 

For self-dual spacetimes our use of the $SL(2,\mathbb{C})_L$ group will be prominent. 
A local $SL(2,\mathbb{C})_L$ null rotation on the spinor dyad $(o^A, \iota^A)$ is given by 
\cite{Chandrasekhar:1985kt, mcintosh-complex-relativity-I, Stephani:2003tm}
\begin{gather}
o^A \rightarrow o^A, \quad \iota^A \rightarrow \iota^A+ b o^A, 
\label{eqn:null-typeI} \\ 
o^A \rightarrow o^A + a \iota^A, \quad \iota^A \rightarrow \iota^A, 
\label{eqn:null-typeII} \\ 
o^A \rightarrow c o^A, \quad \iota^A \rightarrow c^{-1}\iota^A \qquad .
\label{eqn:null-typeIII}
\end{gather} 
These induce corresponding transformations of the Maxwell and Weyl scalars.

{\em Penrose Twistorial double copy.}---A \emph{twistor} is a two-component, complex spinor field $\Omega^A(x)$ satisfying the twistor equation
\be
\nabla_{A'}{}^{(A}\Omega^{B)} = 0,\label{twistor eq1}
\ee
where $\nabla_{AA'}$ is the covariant derivative of the spacetime on which $\Omega^A(x)$ is defined, and spacetime indices are identified with pairs of spinor indices using a soldering form $\sigma_\mu^{AA'}$ in the usual way: $v^{AA'} = \sigma_\mu^{AA'}v^\mu$, and so on.

 Over complexified Minkowski space $\mathbb{CM}$, \eqref{twistor eq1} can be solved exactly to give 
\be
\Omega^A = \omega^A -ix^{AA'}\pi_{A'}, \label{twistorSol}
\nonumber
\ee
where $\omega^A$ and $\pi_{A'}$ are constant spinors. Thus, the space of solutions to the twistor equation---i.e., the twistor space, $\mathbb{T}$---is coordinatized by a pair of spinors $Z = (\omega^A, \pi_{A'})$ and we can regard $\mathbb{T}$ as a four-dimensional complex vector space. The subset of $\mathbb{CM}\,$ defined by the zero locus of an arbitrary twistor
\be
\Omega^A(x) = 0 \label{zero1}
\nonumber
\ee
defines the \emph{incidence relation}
\be
\omega^A = ix^{AA'}\pi_{A'} \, . \nonumber \label{zero2}
\ee

The incidence relation is invariant under complex rescalings $ Z \to  \tau Z$ for any non-zero $\tau \in \mathbb{C}$. This gives rise to the notion of \emph{projective twistor space} $\mathbb{PT}$, the space of twistors up to complex rescalings.

The {\it Penrose transform} \cite{Penrose:1965am, Penrose:1967wn, Penrose:1968me, Penrose:1969ae}
\begin{equation}
\phi_{A_1'A_2'\ldots A_{2s}'}(x)=\frac{1}{2\pi i}\oint_{\Gamma}\pi_{E'}d\pi^{E'}
\pi_{A_1'}\pi_{A_2'} \ldots \pi_{A_{2s}'}[\rho_x g(Z^\alpha)] ,
\label{Penrose}
\end{equation}
relates functions $g(Z^\alpha)$ on $\mathbb{PT}$ with complex spinor fields on spacetime,  a relation further elaborated on below. 
In order for the integrand to be invariant under twistor rescalings, and so defined on $\mathbb{PT}$,
$g(Z^\alpha)$ must be a homogeneous function of the twistor coordinates of degree
$-2(s+1)$. 
Here $\rho_x$ denotes that the incidence relation has been imposed, which imparts the dependence on the
spacetime point $x^{AA'}$ to the field on the left hand side. The integral is performed over any closed contour $\Gamma$ enclosing a single pole, corresponding to a loop on the Riemann Sphere. 

On the Riemann sphere there is a discrete ambiguity in what is meant by the interior and exterior of a closed curve $\Gamma$. The Penrose transform, and more generally the integral of a meromorphic one-form along $\Gamma$, is not ambiguous however, a direct consequence of the ``inside-outside theorem," which asserts that the sum of the residues of any meromorphic one-form on a compact Riemann surface is zero \cite{Griffiths:1994prl}.

The integrand in the Penrose transform is not uniquely determined by the spacetime spinor field. For the Penrose transform relates \emph{equivalence classes} of functions on $\mathbb{PT}$ with complex spinor fields on spacetime, where the equivalence is up to the addition of functions that are holomorphic on one or the other side of the contour $\Gamma$ on the Riemann Sphere. 
That functions holomorphic on the interior of $\Gamma$ do not contribute to the integral is a simple consequence of the residue theorem, while functions holomorphic on the exterior cannot contribute by virtue of the same ``inside-outside theorem." 
\footnote{The data required to specify a spinor field on spacetime taking these redundancies into consideration, including those due to deformations of the contour $\Gamma$,  can be characterized in terms of \v{C}ech cohomology, or equivalently Dolbeault cohomology. See \cite{Huggett:1986fs} chapters 8 and 9 for an exposition. 
}

Remarkably, the field defined by (\ref{Penrose}) solves the zero rest-mass equation $$\nabla^{A_1 A'_1}\phi_{A_1'A_2'\ldots A_{2s}'} = 0$$ for any spin $s$. In particular, the cases $s=0$ and $s =1$ correspond to solutions of the scalar wave equation and vacuum Maxwell equations. The case $s=2$ would be the Bianchi identity if $\nabla^{AA'}$ was a curved-space connection and $\phi_{A'_1A'_2A'_3A'_4}$ was the associated Weyl spinor. However, in the zero rest mass equation  $\nabla^{AA'}$ is a \emph{flat} connection, and $\phi_{A'_1A'_2A'_3A'_4}$ is a \emph{linearized} Weyl spinor. Consequently, the $s=2$ zero rest mass equation is  equivalent to the linearized Bianchi identities because $\Psi_{ABCD}=0$ at zeroeth order.
On the other hand, it can be shown that the $s=2$ zero rest mass equation is also equivalent to the linearized Einstein equations (see e.g. \cite{Wald:1984rg} chapter 13, problem 6).  To be explicit, the fields in the $s=1$ and $s=2$ cases are the
self-dual field strength spinor $\phi_{AB} = \Phi_{AB}$ and the linearized Weyl spinor $\phi_{ABCD} = \Psi_{ABCD}$. Note that for vacuum Kerr-Schild spacetimes, which are solutions of both the full non-linear Einstein equations and the linearized equations, the linearized Weyl scalar happens to equal the full Weyl scalar.

It is often convenient to fix a spinor dyad $(o^A, \iota^B)$, and work with the corresponding Maxwell scalars $$\varphi_0 = \Phi_{AB} o^A o^B, \quad \varphi_1 = \Phi_{AB} o^A \iota^B, \quad \varphi_2 = \Phi_{AB} \iota^A \iota^B,$$ and Weyl scalars
$$ \Psi_0 = \Psi_{ABCD} o^A o^B o^C o^D, \,\, \ldots \,\, ,\Psi_4 = \Psi_{ABCD} \iota^A \iota^B \iota^C \iota^D. \nonumber$$

Motivated by the twistorial double copy introduced in Ref.~\cite{White:2020sfn}, 
we study fields of the form
\begin{align}
\phi_{A_1'A_2'\ldots A_{2s}'}=\frac{\lambda}{2\pi i}\oint_\Gamma dY
\frac{(1,Y)_{A_1'}(1,Y)_{A_2'}\ldots (1,Y)_{A_{2s}'}}{f(Y, u + Y\tilde{\zeta}, \zeta + Yv)^{s+1}} \,.
\label{Pentrans}
\end{align}
Here $\lambda$ is a constant normalization, which is fixed below. 
Compared to Ref.~\cite{White:2020sfn}, we directly link the integrand appearing in the Penrose transform 
to the holomorphic function $f$ on twistor space whose zero set generates exact self-dual vacuum Kerr-Schild spacetimes, described above. 

{\em Self-dual (Kerr)--Taub--NUT.}---The Kleinian self-dual (Kerr)-Taub-NUT solution was recently shown to be diffeomorphic to the non-rotating Kleinian Taub-NUT solution \cite{Crawley_2022}, and subsequently shown to be of the twistorial Kerr-Schild form \cite{Desai:2024fgr, Kim:2024dxo}. It is characterized by the choices
$ \Phi(Y) = i\alpha Y, ~a = b = 1/\sqrt{2}, c = \tilde{c} = 0, $
so that 
\begin{multline*}
f(Y, u + Y \tilde{\zeta},\zeta + Y v) = 
 -\frac{\tilde{\zeta}}{\sqrt{2}}(Y-Y_+)(Y - Y_-), 
\end{multline*}
where 
$Y_\pm = -(\frac{u-v}{\sqrt{2}} + i\alpha \pm \sqrt{(\frac{u - v}{\sqrt{2}} + i\alpha)^2 + 2\zeta\tilde{\zeta}})/{\sqrt{2}\tilde{\zeta}}.$

The zeroth copy obtained from setting $k = 1$ in the Penrose transform evaluates to  
\bea 
\varphi^{PT} &=& 
-\frac{\lambda\sqrt{2}}{\tilde{\zeta}}\frac{1}{Y_+ - Y_-} \, , \nonumber
\eea
where here and in what follows, in evaluating the Penrose transform (\ref{Pentrans}) the contour $\Gamma$ is chosen to encircle the pole $Y=Y_+.$ There is nothing special about this pole, as the other pole $Y=Y_-$ could equally have been chosen, with similar results obtained.

The single copy corresponding to $k = 2$ is found to be 
\begin{gather}
\varphi_0^{PT} =
-\frac{4\lambda}{\tilde{\zeta}^2}\frac{1}{(Y_+ - Y_-)^3}, \quad 
\varphi_1^{PT} =
-\frac{2\lambda}{\tilde{\zeta}^2}\frac{Y_+ + Y_-}{(Y_+ - Y_-)^3}  \nonumber \\
\varphi_2^{PT} =
-\frac{4\lambda}{\tilde{\zeta}^2}\frac{Y_+Y_-}{(Y_+ - Y_-)^3}, \nonumber
\end{gather}
and the double copy $(k = 3)$ results in
\begin{gather}
\Psi_0^{PT} =
-\frac{12\lambda\sqrt{2}}{\tilde{\zeta}^3(Y_+ - Y_-)^5}  , \quad 
\Psi_1^{PT} =
-\frac{6\lambda\sqrt{2}(Y_+ + Y_-)}{\tilde{\zeta}^3(Y_+ - Y_-)^5}  \nonumber \\
\Psi_2^{PT} = 
- \frac{2\lambda \sqrt{2}}{\tilde{\zeta}^3(Y_+ - Y_-)^5}(Y_+^2 + 4Y_+Y_- + Y_-^2)
\nonumber \\
\Psi_3^{PT} = 
- \frac{6\lambda\sqrt{2}Y_+ Y_- (Y_+ + Y_-)}{\tilde{\zeta}^3(Y_+ - Y_-)^5} 
 , \quad 
\Psi_4^{PT} = 
-\frac{12\lambda\sqrt{2}Y_+^2Y_-^2}{\tilde{\zeta}^3(Y_+ - Y_-)^5}.
\nonumber 
\end{gather}
On the Kerr-Schild side, the zeroth copy is 
$\varphi^{KS} = 2P^2V$.
With the vacuum solution $V = m\rho/2P^3$, we find  
$$\varphi^{KS} = \frac{m\rho}{P} \, . $$
In the current example, 
\begin{gather}
\varphi^{KS} = -\frac{m}{\sqrt{\left(\frac{u-v}{\sqrt{2}} + i\alpha\right)^2 + 2\zeta\tilde{\zeta}}} 
= -\frac{m}{\lambda}\varphi^{PT}. \nonumber
\end{gather}
The KS and PT zeroth copies agree once the overall normalization of the Penrose Transform is permanently fixed to be $\lambda=-m$. Turning to 
the single copy, we have
\begin{gather}
\varphi_0^{KS}= 0, \quad 
\varphi_1^{KS} 
= \frac{2m}{\tilde{\zeta}^2}\frac{Y_+ - Y_-}{(Y_+ - Y_-)^3} , \nonumber \\
\varphi_2^{KS} 
= -\frac{4m}{\tilde{\zeta^2}}\frac{1}{(Y_+ - Y_-)^3}. \nonumber 
\end{gather}
For the double copy, from the Kerr-Schild metric we have 
\begin{gather}
\Psi_0^{KS} = \Psi_1^{KS} = 0, \quad 
\Psi_2^{KS} 
= \frac{2\sqrt{2}m}{\tilde{\zeta}^3(Y_+-Y_-)^3} \nonumber \\
\Psi_3^{KS} 
= -\frac{6\sqrt{2}m}{\tilde{\zeta}^3(Y_+-Y_-)^4}, \quad 
\Psi_4^{KS} 
= \frac{12\sqrt{2}m}{\tilde{\zeta}^3(Y_+-Y_-)^5}. \nonumber
\end{gather}
Inspection of the above results reveals that the KS and PT Maxwell and Weyl scalars na\"ively disagree with each other. This situation raises a question: do the Weyl scalars defined by the Penrose Transform describe a different metric? Or in fact, the same one? At this juncture we note that we have not specified the null tetrad implicitly defining the quantities obtained from the Penrose transform. The fact that the KS and PT single and double copies disagree may merely be because the two null tetrads in question are not the same, yet still describe the same physics.

To see that is indeed the resolution of the puzzle, we exhibit null Lorentz transformations that 
transform the two sets of Maxwell and Weyl scalars into each other, which incidentally also  demonstrates that the Penrose transform Weyl scalars implicitly define the same Kerr-Schild metric. 
 
To show the equivalence, perform an $SL(2,\mathbb{C})_L$ transformation (\ref{eqn:null-typeII}) setting $\varphi_0^{PT}$ to zero, 
$$ \varphi_0^{PT} \to   \varphi_0^{PT} + 2b\varphi_1^{PT} + b^2\varphi_2^{PT} \equiv 0,$$
leading to a pair of solutions which, after some algebra, can be expressed as
$b_\pm = -1/Y_\mp.$ 
The remaining Maxwell scalars are transformed according to
\bea 
\varphi_1^{PT} &\to & \varphi_1^{PT} + b_\pm \varphi_2^{PT} \nonumber \\
&=& \pm\frac{2\lambda(Y_+ - Y_-)}{\tilde{\zeta}^2(Y_+ - Y_-)^3}=\pm \frac{\lambda}{m}\varphi_1^{KS} 
\nonumber \\
\varphi_2^{PT}  &\to & \varphi_2^{PT} 
= -\frac{4\lambda Y_+Y_-}{\tilde{\zeta}^2(Y_+-Y_-)^3} \nonumber \\
&=& \frac{\lambda}{m} Y_+Y_-\varphi_2^{KS}. \nonumber
\eea 
Then perform a transformation (\ref{eqn:null-typeIII}) with $c^2 = -Y_+Y_-$. The Maxwell scalars for the Penrose transform and Kerr-Schild sides are then found to be related by the general expression 
\be
\varphi_i^{KS} = \varphi_i^{PT}
\nonumber
\ee 
provided one chooses the root $b_-  = -Y^{-1}_+$ for the $SL(2,\mathbb{C})_L$ transformation parameter of equation (\ref{eqn:null-typeII}). 

To show that the Weyl scalars coming from the Penrose transform and Kerr-Schild construction are equivalent, the same $SL(2,\mathbb{C})_L$
transformations used for the Maxwell scalars are applied, in the same order. Namely, 
\bea 
\Psi_0^{PT}  &\to & \Psi_0^{PT} + 4b_-\Psi_1^{PT} + 6b^2_-\Psi_2^{PT}  
+ 4b^3_-\Psi_3^{PT} + b^4_-\Psi_4^{PT} \nonumber \\
&=&  0  \nonumber \\
\Psi_1^{PT}   &\to& 
\Psi_1^{PT} + 3b_-\Psi_2^{PT} + 3b^2_-\Psi_3^{PT} + b^3_-\Psi_4^{PT} = 0 \nonumber \\
\Psi_2^{PT}   &\to&
\Psi_2^{PT} + 2b_-\Psi_3^{PT} +  b^2_-\Psi_4^{PT}
=  - \frac{2 \lambda \sqrt{2}}{\tilde{\zeta}^3(Y_+ - Y_-)^3} \nonumber \\
&=& - \frac{\lambda}{m} \Psi^{KS}_2 \nonumber \\
\Psi_3^{PT}   &\to&
\Psi_3^{PT} + b_-\Psi_4^{PT} 
=  -\frac{6 \lambda\sqrt{2}Y_+Y_-}{\tilde{\zeta}^3(Y_+ - Y_-)^4}  \nonumber\\
&=& \frac{\lambda}{m} Y_+ Y_- \Psi^{KS}_3 \nonumber\\
\Psi_4^{PT}   &\to& \Psi_4^{PT} =- \frac{12 \lambda\sqrt{2}Y^2_+Y^2_-}{\tilde{\zeta}^3(Y_+ - Y_-)^5} \nonumber \\
&=& - \frac{\lambda}{m} (Y_+ Y_-)^2 \Psi^{KS}_4 \nonumber 
\eea 
where in the above transformations $b_- = -Y^{-1}_+$ is again used, as done for the Maxwell scalars. Then the same choice of transformation (\ref{eqn:null-typeIII}) given by $c^2 = -Y_+Y_-$ leads to the general relation $\Psi_i^{KS} =\Psi_i^{PT}. $

Out of the two roots $b_\pm$ obtained from setting $\varphi^{PT}_0 \to 0$, only one is useful to demonstrating the equivalence of the Penrose and Kerr-Schild Maxwell and Weyl scalars. More generally, the choice of root $b$ is correlated with the choice of which pole the contour $\Gamma$ encloses: $b_\mp$ is found to be correlated with $\Gamma_\pm$. The other root is a valid null transformation, it just isn't useful to that purpose.

This concludes our demonstration that the twistorial and Kerr-Schild double copies for the self-dual (Kerr)-Taub-NUT spacetime are equivalent. We emphasize that while the Penrose transform generically only produces solutions of linearized gravity for $s=2$, we see that when the function on projective twistor space is chosen to be of the Kerr-Schild form, the result is an \emph{exact} solution of the full non-linear vacuum Einstein equations, in this case, the self-dual (Kerr)-Taub-NUT spacetime. The example we have considered is highly symmetric, possessing three additional Killing vectors on top of the one that exists by construction. The solution also enjoys enhanced algebraic symmetry, being of Petrov type D while a generic twistorial Kerr-Schild metric has Petrov type II. Either of these special properties may reasonably lead the reader to question the generality of the equivalence between the Kerr-Schild and Penrose transform double copies within the self-dual sector of the twistorial Kerr-Schild family. We will address these concerns in a companion article that includes explicit examples with Petrov type II, as well as a general proof of the equivalence for self-dual twistorial Kerr-Schild metrics generated by an arbitrary analytic function on projective twistor space.
 
{\em Acknowledgements.}---We thank Kara Farnsworth and Tucker Manton for comments on the manuscript. The work of MG is supported by the 
 U.S. Department of Energy, Office of Science, Office of High Energy physics, under contract number KA2401012 (LANLE83G) at Los Alamos National Laboratory operated by Triad National Security, LLC, for the National Nuclear Security Administration of the U.S. Department of Energy (Contract No. 89233218CNA000001). 
 \newpage

\bibliographystyle{apsrev4-1.bst}
\bibliography{bib}
\end{document}